\documentstyle[psfig,conf_iap,]{article}
\def\kms{km~s$^{-1}$}
\def\degree{$^{\circ}$}

\begin{document}
\heading{%
%
The small-scale structure of the Magellanic Stream
%
} 
\par\medskip\noindent
\author{%
Snezana Stanimirovic$^{1}$, John M. Dickey$^{2}$, Alyson M. Brooks$^{3}$
}
\address{%
Arecibo Observatory, NAIC/Cornell University, HC 3 Box 53995,
Arecibo, Puerto Rico 00612
}
\address{%
Astronomy Department, University of Minnesota, 116 Church Street SE, 
Minneapolis, MN 55455
}
\address{%
Macalester College, Saint Paul, MN 55105
}

\begin{abstract}
We have mapped in neutral hydrogen (HI) two regions at the northern tip of
the Magellanic Stream, known as MS V and MS VI, using the Arecibo
telescope.  The small-scale structure of the MS shows clumpy and head-tail
morphology.  The spatial power spectrum of this  star-free intergalactic
medium has a power-law behavior with the  density slope of $-3.8$. A
gradual steepening of the power-law slope is seen when increasing the
thickness of velocity slices.
 
\end{abstract}

\section{Introduction} 

The Magellanic Stream (MS) is a thin ($\sim$ 10\degree~wide, [2])  tail of
neutral hydrogen (HI), emanating from the Magellanic Clouds and trailing
away for almost  100\degree~on the sky.    This huge HI structure is the
most  fascinating signature of the  wild past interaction of our Galaxy
with the Magellanic Clouds, and the Magellanic Clouds with each other.  We
have undertaken HI observations of two regions, known as MS V and MS VI, at
the northern tip of the MS.  One of the main motivations for this project
is to investigate whether a hierarchy of  structures exists in this  almost
primordial environment, where no  stars have been found yet.

\section{Observations and Data Reduction}
The on-the-fly HI mapping observations were made with the 305 m Arecibo
telescope in June 2000.  Two regions at the northern tip of the MS were
mosaiced with many overlapping maps, which were combined together during
the gridding process.  A linear or quadratic polynomial function, fitted
over emission-free channels, was used for the bandpass removal. The final
angular resolution is 4 arcmin. The noise level is 0.04 K per 1.3 \kms~wide
channel,  which is equivalent to $\sim$ 10$^{17}$ atoms cm$^{-2}$ (rms).
For more information on observations and data reduction see  Stanimirovic
et al., [3].
 
\section{Preliminary Results}

The small-scale structure of the northern tip
of the MS consists of clumps and occasional filaments and/or loop-like 
features. Although there are few prominent large-scale loop-like features,
no systematic appearance of expanding shells of gas is seen in the structure 
of the HI. This supports the `standard' scenario for creation of such features,
whereby shells are the aftermath of stellar winds and supernova explosions 
on the surrounding environment. 
Several clumps have a significant density and 
velocity gradient in the position-velocity diagrams, so called head-tail 
morphology, suggesting that the MS gas is interacting with the Galactic Halo gas.

\begin{figure}
\centerline{\vbox{
\psfig{figure=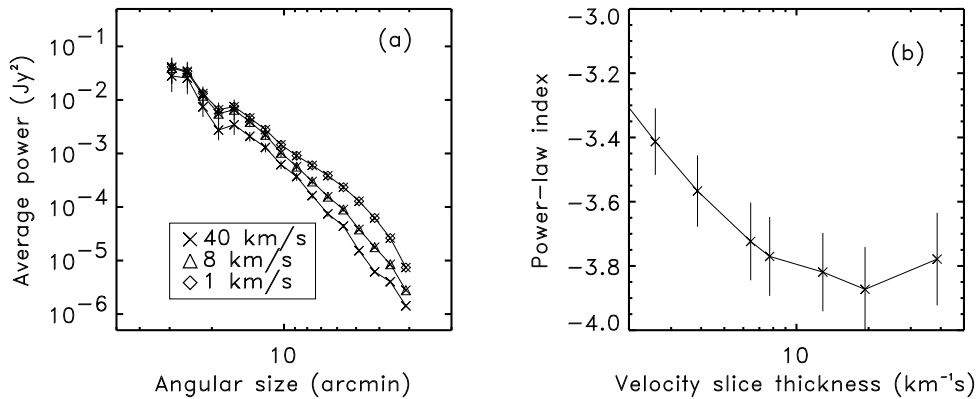,height=6.cm}}}
\caption[]{(a) The 2-D HI spatial power spectrum of intensity fluctuations in 
the MS VI region. Crosses, triangles and diamonds show spectra obtained when
averaging 30, 6 or 1 velocity channels, respectively.
(b) A variation of the 2-D HI power spectrum slope with velocity slice
thickness. A gradual steepening from $-3.1$ to $-3.8$ is seen showing the
transition from the velocity to the density dominated regime.} 

\end{figure}
 
To investigate hierarchy of structures we have derived the 2-D spatial
power spectra for the MS VI region. These spectra can be fitted by a
power-law both in the case of individual channels and the integrated column
density distribution, as shown in Fig. 1, panel (a). The power-law slope
gradually steepens, from $-3.1$ to $-3.8$, when changing thickness of
velocity slices from 1.3 \kms~to ~40 \kms, see Fig. 1 panel (b). This is in
an excellent agreement with the turbulent theory by Lazarian \& Pogosyan,
[1]. From the thickest velocity slices, the 3-D density power spectrum is
$-3.8 \pm 0.1$, which is remarkably close to the Kolmogorov spectrum.

\vfill
\end{document}